\documentstyle[aps,preprint]{revtex}
\begin{document}
\draft
\title{Image resolution depending on slab thickness and object distance in a two-dimensional photonic-crystal-based superlens}
\author{Xiangdong Zhang}
\address{Department of Physics, Beijing Normal University, Beijing
100875, P. R. China}

\maketitle

\begin{abstract}
Based on the exact numerical simulation and physical analysis, we
have demonstrated all-angle single-beam left-handed behavior and
superlens for both TE and TM modes in a two-dimensional coated
photonic crystals. The imaging behaviors by two-dimensional
photonic-crystal-based superlens have been investigated
systematically. Good-quality images and focusing, with relative
refractive index of -1, have been observed in these systems for
both polarized waves. In contrast to the images in near-field
region for the lowest valence band, non-near-field images,
explicitly following the well-known wave-beam negative refraction
law, have been demonstrated. The absorption and compensation for
the losses by introducing optical gain in these systems have also
been discussed. Thus, extensive applications of such a phenomenon
to optical devices are anticipated.

\end{abstract}
\narrowtext

\section{INTRODUCTION}

Recently there has been a great deal of interest in studying a
novel class of media that has become known as the left-handed
materials (LHMs)[1-20]. These materials are characterized by
simultaneous negative permittivity $\epsilon$ and negative
permeability $\mu$. Properties of such materials were analyzed
theoretically by Veselago over 30 years ago[1]. As was shown by
veselago, the LHMs possess some peculiar electromagnetic
properties such as inverse Snell's law, reversed Doppler shift,
and reversed Cherenkov radiation. It had also been suggested that
slab of the LHM could be employed as an unconventional flat lens.

Due to the absence of naturally occurring materials having both
negative $\epsilon$ and negative $\mu$, Veselago's prediction did
not receive much attention until recently, when a system
consisting an array of split-ring resonators and metallic wires
was prepared and demonstrated to have negative refractive index
experimentally [3,4]. Subsequently, some physical properties of
the LHMs were analyzed by many authors [5-20]. Pendry [5]
predicted that the LHM slab can amplify the evanescent waves and
the flat lens constructed from such a material with
$\epsilon=\mu=-1$ could in principle work as ``perfect" lens
(superlens). Although the concept of superlens was questioned by a
number of authors [21], more detailed physical considerations had
shown that the construction of ``almost perfect" lens is indeed
possible [22-27]. Recently, such image behaviors have been
observed by some numerical simulations [22-24] and experimental
measurements [24-27]. However, only near-field images were
demonstrated and extensive applications of such a phenomenon were
limited [22-27].

 It was shown that the negative refraction could also occur in
photonic crystal (PC)[28-39]. The physical principles that allow
negative refraction in them arise from the dispersion
characteristics of wave propagation in a periodic medium, which
can be well described by analyzing the equifrequency surface (EFS)
of the band structures [28-37]. In the PC structures, there are
two kinds of cases for negative refraction occuring [31]. The
first is the left-handed behavior as being described above[29-32].
In this case, $\vec{k}$, $\vec{E}$ and $\vec{H}$ form a
left-handed set of vectors (i.e., $\vec{S}\cdot\vec{k}<0$, where
$\vec{S}$ is the Poynting vectors). Another case is that the
negative refraction can be realized without employing a negative
index or a backward wave effect[33-37]. In this case, the PC is
behaving much like a uniform right-handed medium(i.e.
$\vec{S}\cdot\vec{k}>0$). Recently, Luo $et$ $al.$[33] have shown
that all-angle negative refraction could be achieved at the lowest
band of two-dimensional(2D) PC in the case of
$\vec{S}\cdot\vec{k}>0$. The advantages of the negative refraction
in the lowest valence band are single-mode and high transmission.
These can help us to design microsuperlens and realize the
focusing of the wave. Very recently, the subwavelength focusing
and image by 2D PC slab have been observed experimentally [34-36].
Absolute negative refraction and imaging of unpolarized
electromagnetic wave by 2D PC slabs have also been obtained [37].
However, due to the anisotropy of dispersion in 2D PC, such images
only appear in near-field region[38,39].

How to realize a good-quality non-near-field image becomes an
important issue. The prerequisite condition to realize such a
phenomenon is the negative refraction which possesses the
single-mode and high transmission. Although some works[29] have
shown that the left-hand behavior and focusing exist in the 2D PC,
the multiple-mode and low transmission in high frequencies affect
the features of focusing[31,32]. It is well known that the
electromagnetic (EM) wave can be decomposed into TM modes (S wave)
and TE modes (P wave) for the 2D PC structures [40]. However, the
investigations[29] have shown that air-hole-type 2D PCs possess
good left-handed behavior for TM modes, and a pillar type 2D PC
prefers TE modes. In this paper, we will demonstrate that coated
cylinder PCs with triangular lattice posses double features. They
have not only good left-handed behavior for the TM modes, but also
for the TE modes. Most interestingly, the all-angle single-beam
left-handed behaviors with relative refractive index of -1 for
both polarized waves have been found in these systems. Thus,
high-quality focusing and imaging behavior have been obtained. In
contrast to the images in the near-field region for the lowest
valence band, non-near-field images, explicitly following the
well-known wave-beam negative refraction law, have been
demonstrated. The absorption and compensation for the losses by
introducing optical gain in these systems have also been
discussed.

The rest of this paper is arranged as follows. In Sec. II, we
demonstrate the all-angle single-beam left-handed behaviors for
both TE and TM modes in a two-dimensional coated photonic
crystals. The good-quality non-near-field imaging behaviors are
discussed in Sec. III. In Sec. IV, we analyze the effect of
absorption and gain. The conclusions are given in Sec. V.

\section{LEFT-HANDED BEHAVIOR in 2D COATED CYLINDER PC}

We consider a 2D triangular lattice of coated cylinders immersed
in a air background with lattice constant $a$. The coated
cylinders have metallic cores coated with a dielectric coating.
The radii of metallic core and coated cylinder are 0.25a and
0.45a, respectively. The dielectric constants of dielectric
coating are taken as $11.4$ for the S wave and $7.0$ for the P
wave. For the metallic component, we use the frequency-dependent
dielectric constant [41],
\begin{equation}
\epsilon=1-\frac{f_{p}^2}{f(f+i\gamma)},
\end{equation}
where $f_{p}$ and $\gamma$ are the plasma frequency and the
absorption coefficient. Following Ref.41, for all numerical
calculations carried out in this work, we have chosen
$f_{p}=3600$THZ and $\gamma=340$THZ, which corresponds to a
conductivity close to that of Ti. However, our discussion and
conclusions given below can apply to other metal parameters as
well. In order to simplify the problem, we first consider the
cases without absorption ($\gamma=0$). The effect of absorption
will be discussed at the latter part.

For all calculations throughout this paper, we adopt the
multiple-scattering Korringa-Kohn-Rostoker method [42] as our main
computational tool, both to calculate the photonic band structure
in the reciprocal space and to perform numerical simulations for
wave propagating in the finite real space. The multiple-scattering
 method is not only success in the calculations of band structure,
it is also best suited for a finite collection of cylinders with a
continuous incident wave of fixed frequency. For circular
cylinders, the scattering property of the individual cylinder can
be obtained analytically, relating the scattered fields to the
incident fields. The total field, which includes the incident plus
the multiple-scattered field, can then be obtained by solving a
linear system of equations, whose size is proportional to the
number of cylinders in the system. Both near field and far field
radiation patterns can be obtained straightforwardly. So, such a
method is a very efficient way of handling the scattering problem
of a finite sample containing cylinders of circular cross
sections, and it is capable of reproducing accurately the
experimental transmission data, which should be regarded as exact
numerical simulation. The detailed description of this method has
been given in Ref. [42].

  The calculated results of band structure for the S wave and the P
wave are plotted in the Fig.1(a) and (b), respectively. We focus
on the problems of wave propagation in the second bands marked by
dotted lines in Fig.1(a) and (b). Owing to the strong scattering
effects, it is generally difficult to describe the propagation
behavior of EM wave in the PC in a simple yet accurate way.
However, a lot of theoretical and experimental practices[28-39]
have shown that the overall behavior of the wave propagation
within a PC can be well described by analyzing the equifrequency
surface (EFS) of the band structures, because the gradient vectors
of constant-frequency contours in k-space give the group
velocities of the photonic modes. Thus, the propagation direction
of energy velocity of EM wave can be deduced from them. The EFS
contours of the above system for the S wave and the P wave at
several relevant frequencies are demonstrated in Fig.2(a) and (b),
respectively.

 It is clear from the figures that some EFS contours such as
$\omega=0.42-0.47(2\pi c/a)$ for the S wave and the P wave are
very close to a perfect circle, indicating that the crystal can be
regarded as an effective homogeneous medium at these frequencies.
At the same time, we also notice that the frequencies increase
inwards for both cases, meaning that $\vec{S}\cdot\vec{k_{i}}<0$
and the group velocities ($v_{g}$) are opposite to the phase
velocity. Here, $\vec{S}$ and $\vec{k_{i}}$ represent the Poynting
vector and wave vector, respectively. These indicate that the
transmitting features of the wave in the above PC structures are
the left-handed behavior. The conservation of the surface-parallel
wave vector would result in the negative refraction effect in
these cases, which the direction of the refracted wave inside the
PC can also estimated from the EFS. Then, we apply Snell's law in
these cases, the negative indexes of refraction can also be
obtained. Fig.3(a) and (b) show the negative effective indexes as
function of frequencies for the S wave and the P wave,
respectively. It is interesting that the cases with relative
refractive index of -1 for the S wave at $\omega=0.42(2\pi c/a)$
and the P wave at $\omega=0.43(2\pi c/a)$ have been found.

In order to test the above analysis, we do numerical simulations
in the present systems. We take the slab samples which consist of
13-layer coated cylinders in the air background with a triangular
arrays. The surface normal of the PC slab is along $\Gamma K$
direction. The parameters of coated cylinders are the same to the
cases in Fig.1.  When a slit beam with a half width $2a$ goes
through the slab material, it will be refracted two times by two
interfaces of the slab. The shape of the sample and a snapshot of
the refracted process are shown on the top of Fig.4. There are
different ray-traces for the wave transmitting through the slab
sample with various effective indexes, when the wave does not
incident on the interface with normal direction. From the
ray-traces, we can deduce the effective refraction index of the
slab material.

The simulations are based on a highly efficient and accurate
multiple-scattering method [42]. In our calculations, the widths
of the samples are taken enough large, such as 40a, to avoid the
edge diffraction effects. The calculated results for the S wave
and the P wave are plotted in Fig.4(a) and (b), respectively. The
field energy patterns of incidence and refraction are shown in the
figures. The arrows and texts illustrate the various beam
directions. The geometries of the slab are also displayed. It can
be clearly seen that the energy fluxes of refraction wave travel
following the negative refraction law for both polarized waves.
The negative refraction indexes for the S wave and the P wave
obtained from the exact numerical simulation are marked as dark
dot in Fig.3(a) and (b), respectively. Comparing them with the
estimated results (solid lines in Fig.3(a) and (b)) from the EFS ,
we find that the agreements between them are well around the
region of effective refractive index of -1 for both polarized
waves. When the effective refractive indexes deviate largely from
-1, some differences can be found. This is due to the Goos-Hanchen
effect, which had been discussed in Ref.[43].

Varying the angle of inclination of the sample, we have checked
the cases with various incident angles.  The calculated and
analytical results of refracted angle $\theta$ versus incident
angle $\theta_0$ at $\omega=0.42(2\pi c/a)$ for the S wave and
$\omega=0.43(2\pi c/a)$ for the P wave are summarized in Fig.5 by
the circle  and triangular dark dots, respectively. Because the
frequencies are below 0.5$(2\pi c/a)$, all-angle single-beam
negative refractions have been observed at these frequencies. More
interestingly, $\theta$ is linearly proportional to $\theta_0$ in
the whole angle region for both polarized waves. This feature is
close to the ideal LHM system that can serve as a perfect
superlens [5].

\section{IMAGE DEPENDING ON SLAB THICKNESS AND OBJECT DISTANCE}

  It is well known that an important application of negative refraction
materials is the microsuperlens [5]. Ideally, such a superlens can
focus a point source on one side of the lens into a real point
image on the other side even for the case of a parallel sided slab
of material. It possesses some advantages over conventional
lenses. For example, it can break through the traditional
limitation on lens performance and focus light on to an area
smaller than a square wavelength.

 In order to model such a superlens, we
take a slab sample with 40a width and 11a thickness. A
continuous-wave point source is placed at a distance 5.5a (half
thickness of the sample) from the left surface of the slab. We
first discuss the case for the point source of the S wave. The
frequency of the incident wave emitting from such a point source
is $0.42(2\pi c/a)$, which is corresponded to the case with
relative refractive index of -1. If the wave transmits in such a
2D PC slab according to the well-known wave-beam refraction law,
one should observe the focusing point in the middle of the slab
and the image at the symmetric position in the opposite side of
the slab, as being depicted by the simple picture on the top of
Fig.6.

To see whether or not such a phenomenon exists, we employ the
multiple-scattering method [42] to calculate the propagation of
waves in such a system. A typical field intensity pattern for the
S wave across the above slab sample is plotted in Fig. 6(b).
 X and Y present vertical and transverse direction of wave propagating,
respectively. The field intensity in figure is over $30a\times
30a$ region around the center of the sample. The geometry of the
PC slab is also displayed for clarity of view. The high quality
image in the opposite side of the slab and the focusing in the
middle of the slab according to the wave-beam refraction law are
observed clearly. A closer look at the data reveals a transverse
size (full size at half maximum) of the image spot as $0.7a$ (or
$0.3\lambda$), which is well below the conventional diffraction
limit.

In order to clarify the sample thickness dependence of the image
and focusing, we have also checked a series of slab samples with
various thickness.  Similar phenomena have also been observed. For
example, Fig. 6(a) shows the calculated field energy pattern for a
7a thick sample. A monochromatic point source with
$\omega=0.42(2\pi c/a)$ is placed at a distance of half thickness
of the sample (3.5a) from the left surface of the slab and its
image is found again near the symmetric position in the opposite
side of the slab.

To have a more complete vision on the imaging effect of this type
of superlens, we move the light source and see what happens to the
imaging behavior. We first put a point source near the left
surface of the slab. In this case, the refracted process following
wave-beam negative refraction law is shown on the top of Fig.7.
The calculated intensity distribution is plotted in Fig.7. In our
simulations, the point source is placed at a distance of 2.0a from
the left surface of the sample and the image is found near 9a from
the right surface. The excellent agreements between the
simulations and the estimated results from the rules of geometric
optics are obvious. The corresponded result that a point source is
far from the left surface is displayed in Fig.8. In this case, the
point source is placed at a distance of 9.0a from the left surface
of the sample and the image is found near 2a from the right
surface. Comparing the simulations with the snapshot of refracted
process according to the wave-beam negative refraction law shown
on the top of Fig.8, we again find the excellent agreements
between them.

The above results are only for the S wave, in the following, we
will investigate the case of the P wave. We take the PC slab
samples with 40a width and various thicknesses. A point source of
the P wave is placed at a distance of half thickness of the sample
from the left surface of the slab.  The frequency of the incident
wave emitting from such a point source is $0.43(2\pi c/a)$, which
is corresponded to the case with relative refractive index of -1
for the P wave. The propagation behaviors of the P waves in such
systems are still calculated by the multiple-scattering method
[42]. Fig.9(a) and (b) show the cases with 7a and 13a thick slab,
respectively. Similar features to Fig.7 for the S wave are found.
If we move the source position, the effect of source position on
the image for the P wave can also be checked. The calculated
results under two kinds of source position for the P wave are
plotted in Fig.(10)(a) and (b), respectively. The cases for the P
waves are similar to those of the S waves in Fig.8.

These observations indicate clearly that the imaging behaviors
depend on the slab thickness and the object distance, explicitly
following the well-known wave-beam negative refraction law.
Therefore, such PC slabs in some frequencies can be considered as
homogeneous effective medium with effective refraction index of
-1. The high-quality focusing and images can be realized in these
PC systems for both S wave and P wave.

\section{EFFECT OF ABSORPTION AND GAIN}

The above investigations have shown that the PC slab consisting of
coated cylinders is actually considered as a good superlens for
both polarized waves.  The common features of these coated systems
are that they all include metal components. Therefore, the
absorption for these systems is inevitable. Lou $et$ $al.$ [38]
have pointed that the central image peak disappear and the image
degrade gradually with the increase of absorption. However,
fortunately, the loss can be overcome by introducing the optical
gain in the systems. Recently, Ramakrishna and Pendry [19] have
suggested a method to remove the absorption by introducing optical
gain into the lens made from a multilayers stack of thin
alternating layers of silver and dielectric medium. Here, we
borrow their idea and introduce the optical gain in the 2D PC
superlens.

 Fig.11(a) shows the intensity distribution as a function of
transverse coordinate ($y/a$) for the S wave at the image plane
(5.5a away from the second interface). Curve {\it A} is
corresponded to the perfect case without absorption and curve {\it
B} to that with absorption, in this case $\gamma$ is taken as
$340$THZ. Comparing the curve {\it A} with the curve {\it B}, we
find that the central peak of image decrease with the introducing
of absorption, which is agree with the analysis of Ref.[38]. This
is also consistent with the numerical studies of left-handed
structures constructed from split-ring resonators in Ref.[7,8].
The result by introducing gain to remove the absorption for the
corresponding case is plotted in Fig.11(b). Curves {\it B} in the
Fig.11(a) and (b) are the same one, and curve {\it C} in the
Fig.11(b) is the result with the dielectric constant
$\epsilon=11.4-0.08i$ for the dielectric part of coated cylinder.
We do not find any difference between curve {\it A} in the
Fig.11(a) and curve {\it C} in the Fig.11(b).

  Similar phenomenon can also be found for the P wave. Fig.12(a) and
(b) show the corresponded case of the P wave. Fig.12(a) represents
the intensity distribution as a function of transverse coordinate
($y/a$) for the P wave at the image plane (5.5a away from the
second interface). Comparing the curve {\it A} without absorption
 with the curve {\it B} in presence of absorption in Fig.12(a), we
find that the absorption decreases the central peak of the image
as the case of the S wave. However, with the introducing of gain
such as $\epsilon=11.4-0.06i$ for the dielectric part of coated
cylinder, the lose due to absorption can be compensated
completely. Curve C in Fig.12(b) represents such a case. In fact,
for any cases of absorption, the losses can always be compensated
by introducing fitted gain for both polarized waves. Thus, the
lens based on the above 2D PC can work well even in presence of
absorption.

\section{SUMMARY}

Through the exact numerical simulation and physical analysis, we
have demonstrated all-angle single-beam left-handed behavior for
both TE and TM modes in the 2D coated photonic crystals. More
interestingly, the relative refractive index of -1 for both
polarized waves have been found. Furthermore, the refracted angle
is linearly proportional to the incident angle in the whole angle
region for both polarized waves have been demonstrated. These
feature are close to the ideal LHM system that can serve as a
perfect superlens. The imaging behaviors by 2D coated
photonic-crystal-based superlens have been investigated
systematically. Good-quality images and focusing, with relative
refractive index of -1 and explicitly following the well-known
wave-beam negative refraction law, have been observed in these
systems for both polarized waves.

Our above results are in contrast to the previous investigations
about the image behaviors for the LHM-based superlenes and the
PC-based microsuperlenes at the lowest valence band. For these
cases, the images only appear in the near-field region, which does
not follow rules of geometric optics [22-27,38,39]. According to
Pendry's analysis [5] , the ``perfect" image arises from the
enhancement of evanescent components of incoming waves, and
surface plasmons play an important role in the ``perfect" imaging
[5,21]. So, the quality of the image is affected by many factors
such as interface feature, finite-size and cavity resonance
[22-27]. These will limit further applications. Here, our
superlens based on the coated PC systems with triangular lattice
can overcome these restrictions. Thus, extensive applications of
such a phenomenon to optical devices are anticipated.

 In addition, the absorption and compensation for the losses by introducing
optical gain in these systems have also been discussed. In
general, for the PC structures with metal components, increased
absorption in metals prohibits the scaling of these structures to
the optical wavelengths. However, since the losses by absorption
can always be compensated by introducing fitted gain in our coated
PC systems, many negative refraction phenomena that have been
observed in the microwave regime can also be found in the optical
wavelengths. These features make the PC slabs consisting of coated
cylinders promising for application in a range of optical devices,
such as a superlens for visible light.

\begin{center}
Acknowledgments
\end{center}
This work was supported by the National Natural Science Foundation
of China (Grant No.10374009) and the National Key Basic Research
Special Foundation of China under Grant No.2001CB610402. The
project sponsored by SRF for ROCS, SEM and the Grant from Beijing
Normal University.

\newpage

FIGURE CAPTIONS

Fig.1, The calculated photonic band structures of a triangular
lattice of coated cylinder in air for S wave (a) and P wave (b).
The radii of the dielectric cylinder and inner metallic cylinder
are $R=0.45a$ and $r=0.25a$, respectively. The dielectric
constants are $\epsilon=11.4$ for the S wave and $\epsilon=7$ for
the P wave. Dotted lines mark the region for negative refraction.

Fig.2,Several constant-frequency contours for S wave (a) and P
wave (b) of the second band of the 2D PCs which are corresponded
to the cases in Fig.1(a) and (b), respectively. The numbers in the
figure mark the frequencies in unit of $2\pi c/a$. $k_{i}$ and
$v_{g}$ represent the wave vector and the group velocity,
respectively.

Fig.3, Effective indexes versus frequencies for S wave (a) and P
wave (b). The crystals and parameters are identical to those in
Fig.1.  The positions of relative refractive index of -1 are
marked by dotted lines.

Fig.4, Simulation of negative refraction. The shape of the sample
and a snapshot of refraction process are shown on top of the
figure.  The intensities of electric field for S wave (a) and
magnetic field for P wave (b) for incidence and refraction are
shown. The 2D PC slabs with 13-layer are marked as dark dots in
figures. The frequencies of incident wave are $\omega=0.42(2\pi
c/a)$ for the S wave and $\omega=0.43(2\pi c/a)$ for the P wave.
The crystals and parameters in (a) and (b) are corresponded to
those in Fig.1(a) and (b), respectively.

Fig.5, The angles of refraction ($\theta$) versus angles of
incidence ($\theta_0$) at $\omega=0.42(2\pi c/a)$ for S wave and
$\omega=0.43(2\pi c/a)$ for P wave. Circle dots are corresponded
to the S wave and triangular dots to the P wave.

Fig.6,  (a) The intensity distributions of point source and its
image across a 7a 2D PC slab  at frequency $\omega=0.42(2\pi c/a)$
for S wave. (b) The corresponding case for a slab with 11a
thickness. Schematic picture depicting the lensing of a source by
a PC slab to an image are shown on top of the figure.

Fig.7,  The intensity distributions of point source and its image
across a 11a 2D PC slab at frequency $\omega=0.42(2\pi c/a)$ for S
wave. The point source is placed at 2a distance from the left
surface of the slab. Schematic picture depicting the lensing of a
point source by a PC slab to an image are shown on top of the
figure.

Fig.8,  The intensity distributions of point source and its image
across a 11a 2D PC slab at frequency $\omega=0.42(2\pi c/a)$ for S
wave. The point source is placed at 9a distance from the left
surface of the slab. Schematic picture depicting the lensing of a
point source by a PC slab to an image are shown on top of the
figure.

Fig.9,  (a) The intensity distribution of point source and its
image across a 7a 2D PC slab  at frequency $\omega=0.43(2\pi c/a)$
for P wave. (b) The corresponding case for a slab with 11a
thickness. The point source is placed at a distance with half
thickness of the sample from the left surface of the slab.

Fig.10,  The intensity distribution of point source and its image
across a 11a 2D PC slab at frequency $\omega=0.43(2\pi c/a)$ for P
wave. (a) and (b) represent the cases with different source
positions, 2a and 9a distances from the left surface of the slab,
respectively.

Fig.11,  Intensity distribution along the transverse (y) direction
at the image plane for S wave. (a) The case with absorption (B)
and that without absorption (A). (b) The case with absorption (B)
and that with absorption and gain (C). The crystal and parameters
are identical to those in Fig.6.

Fig.12,  Intensity distribution along the transverse (y) direction
at the image plane for P wave. (a) The case with absorption (B)
and that without absorption (A). (b) The case with absorption (B)
and that with absorption and gain (C). The crystal and parameters
are identical to those in Fig.9.


\newpage
\begin{references}
\bibitem{1}V.G.Veselago, Sov.Phys.Uspekhi {\bf 8}, 2854 (1967)[Sov. Phys. Usp. 10, 509(1968)].
\bibitem{2}J. B. Pendry, A.J.Holden, D.J.Robbins, and W.J.Stewart, IEEE Trans. Microwave Theory
Tech. {\bf 47}, 2075 (1999).
\bibitem{3}D.R.Smith, W.J.Padilla, D.C.View, S.C.Nemat-Nasser, and S.Schultz,
Phys. Rev. Lett. {\bf 84}, 4184 (2000); D.R.Smith and N. Kroll,
Phys. Rev. Lett. {\bf 84}, 2933 (2000).
\bibitem{4}R. A. Shelby, D.R.Smith, and S.Schultz, Science {\bf 292}, 77 (2001).
\bibitem{5}J. B. Pendry, Phys. Rev. Lett. {\bf 85}, 3966 (2000).
\bibitem{6}P.Markos and C.M.Soukoulis, Phys. Rev. E {\bf 65}, 036622(2002); Phys. Rev. B {\bf 65}, 033401(2002).
\bibitem{7} Markos, I. Rousochatzakis and C. M. Soukoulis, Phys. Rev. E 66 045601(R) (2002).
\bibitem{8} R.B. Greegor, C.G.Parazzoli, K.Li and M.H.Tanielian, Appl. Phys. Lett. 82, 2356(2003).
\bibitem{9}S. Foteinopoulou, E.N. Economou and C.M.Soukoulis,
Phys. Rev. Lett. {\bf 90}, 107402(2003).
\bibitem{10}J. Pacheco, Jr., T.M.Grzegorczyk, T B.I.Wu, Y.Zhang and J.A.Kong,
Phys. Rev. Lett. {\bf 89}, 257401(2002).
\bibitem{11}Y. Zhang, B. Fluegel, and  A. Mascarenhas, Phys. Rev. Lett. {\bf 91}, 157404(2003).
\bibitem{12}D.R.Smith and D.Schurig, Phys. Rev. Lett. {\bf 90}, 077405 (2003).
\bibitem{13}A.A.Houck, J. B. Brock and I. L. Chuang,
Phys. Rev. Lett. {\bf 90}, 137401(2003); C.G.Parazzoli, R. B.
Greegor, K. Li, B. E. C. Koltenbah, and M. Tanielian, Phys. Rev.
Lett. {\bf 90}, 107401 (2003).
\bibitem{14}J. Li,  Lei Zhou, C.T.Chan and P.Sheng,
Phys. Rev. Lett. {\bf 90}, 083901(2003).
\bibitem{15}R. Ziokowski and E. Heyman, Phys. Rev. E {\bf 64}, 056625(2001).
\bibitem{16}Focus Issue ``Negative refraction and metamaterials", Optics Express {\bf vol. 11},no.7 (2003).
\bibitem{17}G. Shvets, Phys. Rev. B {\bf 67},035109(2003).
\bibitem{18}V. A. Podolskiy, A.K.Sarychev and V.M. Shalaev, Optics Express {\bf 11},no.735 (2003).
\bibitem{19}S. A. Ramakrishna and J.B. Pendry, Phys. Rev. B {\bf 67}, 201101(R)(2003).
\bibitem{20} A.A.Zharov, I.V.Shadrivov and Y. S. Kivshar, Phys. Rev. Lett. {\bf 91},
037401(2003); V. M. Agranovich, Y.R.Shen, R.H.Baughman and A.A.
Zakhidov, Phys. Rev. B {\bf 69}, 165112 (2004).
\bibitem{21}G.W.'t Hooft, Phys. Rev. Lett. {\bf 87}, 249701(2001);
J. M. Williams, Phys. Rev. Lett. {\bf 87}, 249703(2001); N.Garcia
and M.Nieto-Vesperinas, Phys. Rev. Lett. {\bf 88}, 207403(2002);
A.L.Pokrovsky and A. L. Efros,  Phys. Rev. Lett. {\bf 89}, 093901
(2002).
\bibitem{22}R.Merlin, Appl.Phys.Lett. {\bf 84}, 1290(2004).
\bibitem{23}L. Chen, S. He and L. Shen, Phys. Rev. Lett. {\bf 92}, 107404 (2004).
\bibitem{24} D.R. Smith, D. Schurig, J. J. Mock, P. Kolinko, and P. Rye, Appl. Phys. Lett. 84, 2244 (2004).
\bibitem{25} Zhaowei Liu, Nicholas Fang, Ta-Jen Yen, and Xiang Zhang, Appl.Phys.Lett. {\bf 83}, 5184(2003).
\bibitem{26}A. Grbic and G.V.Eleftheriades, Phys. Rev. Lett. {\bf 92}, 117403 (2004).
\bibitem{27}A.N.Lagarkov and V.N.Kissel, Phys. Rev. Lett. {\bf 92}, 07701 (2004).

\bibitem{28}H. Kosaka, T. Kawashima, A. Tomita, M.Notomi, T.Tamamura, T.Sato, and S.Kawakami,
Phys. Rev. B {\bf 58}, 10096(1998).
\bibitem{29}M.Notomi, Phys. Rev. B {\bf 62}, 10696(2000).
\bibitem{30}B.Gralak, S.Enoch, and G.Tayeb, J. Opt. Soc. Am. A {\bf 17}, 1012(2000).
\bibitem{31}S. Foteinopoulou and C.M.Soukoulis,
Phys. Rev. B {\bf 67}, 235107(2003).
\bibitem{32}P.V.Parimi, W.T.Lu, P.Vodo, J.Sokoloff, J.S. Derov and S. Sridhar, Phys. Rev. Lett. {\bf 92}, 127401 (2004).

\bibitem{33}C.Luo, S.G.Johnson, J.D.Joannopoulos, J.B.Pendry, Phys. Rev. B {\bf 65},
201104(R)(2002); optics express, {\bf 11}, 746(2003); C.Luo,
S.G.Johnson, J.D.Joannopoulos, Appl.Phys.Lett. {\bf 83},
2352(2002).
\bibitem{34}E. Cubukcu, K. Aydin, E. Ozbay, S. Foteinopoulou and
C.M.Soukoulis, Nature (London) {\bf 423}, 604(2003).
\bibitem{35} P. V. Parimi, W.T.Lu, P. Vodo and S. Sridhar, Nature {\bf 426}, 404(2003).
\bibitem{36}E. Cubukcu, K. Aydin, E. Ozbay, S. Foteinopoulou and
C.M.Soukoulis, Phys. Rev. Lett. {\bf 91}, 207401(2003).
\bibitem{37} Xiangdong Zhang, phys. Rev. B (accepted).
\bibitem{38}C.Luo, S.G.Johnson, J.D.Joannopoulos, J.B.Pendry, Phys. Rev. B {\bf 68},
045115(2003).
\bibitem{39}Z.Y. Li and L.L. Lin, Phys. Rev. B {\bf 68},245110(2003).
\bibitem{40}J.D.Joannopolous, R.D.Meade, and J.N.Winn, Photonic Crystals (Princeton University, Princeton, 1995).
\bibitem{41}M. M. Sigalas, C. T. Chan, K. M. Ho and C.M. Soukoulis, phys. Rev. B {\bf 52}, 11744
(1995); L.M. Li, Z.Q. Zhang and X. Zhang, phys. Rev. B {\bf 58},
15589 (1998).
\bibitem{42}L.M.Li and Z.Q.zhang,phys. Rev. B {\bf 58}, 9587
(1998); X.Zhang, Z.Q.Zhang, and L.M.Li, C. Jin, D. Zhang, B. Man
and B. Cheng, Phys. Rev. B {\bf 61}, 1892 (2000).
\bibitem{43}  D. Felbacq and R. Smaali , Phys. Rev. Lett. {\bf 92}, 193902(2004).











\end{references}
\end{document}